\documentclass[letterpaper, 10 pt, conference]{ieeeconf}  

\IEEEoverridecommandlockouts                              

\usepackage{flushend}

\usepackage{enumerate}
\usepackage{color}
\usepackage{graphicx}
\usepackage{cite}
\usepackage{pifont}

\graphicspath{{images/}}

\newcommand{\MATE}{{\itshape MATE}}

\title{\LARGE \bf Methodological Approach for the Evaluation\\of an Adaptive and Assistive Human-Machine System}

\author{Lorenzo Sabattini$^{1}$, Valeria Villani$^{1}$, Julia N. Czerniak$^{2}$, Frieder Loch$^{3}$,\\Alexander Mertens$^{2}$, Birgit Vogel-Heuser$^{3}$ and Cesare Fantuzzi$^{1}$
\thanks{$^{1}$L. Sabattini, V. Villani, and C. Fantuzzi are with the Department of Sciences and Methods for Engineering (DISMI), University of Modena and Reggio Emilia, Reggio Emilia, Italy
        {\tt\small \{lorenzo.sabattini, valeria.villani,  cesare.fantuzzi\}@unimore.it}}
\thanks{$^{2}$Institute of Industrial Engineering and Ergonomics, RWTH Aachen University, Aachen, Germany
    	{\tt\small \{j.czerniak, a.mertens\}@@iaw.rwth-aachen.de}}
\thanks{$^{3}$F. Loch and B. Vogel-Heuser are with the Institute of Automation and Information Systems, Technical University of Munich, Munich, Germany
        {\tt\small \{frieder.loch, vogel-heuser\}@tum.de}}
\thanks{This work has been supported by the INCLUSIVE collaborative project, which has received funding from the European Union's Horizon 2020 Research and Innovation Programme under grant agreement No 723373.}%
}

\begin{document}

\maketitle
\thispagestyle{empty}
\pagestyle{empty}

\begin{abstract}
With the increasing complexity of modern industrial automatic and robotic systems, an increasing burden is put on the operators, who are requested to supervise and interact with such complex systems, typically under challenging and stressful conditions. To overcome this issue, it is necessary to adopt a responsible approach based on the anthropocentric design methodology, such that machines adapt to the humans capabilities. Moving along these lines, a methodological approach called \MATE{} was introduced in~\cite{Villani_2018_RAM}, which consists in devising complex automatic or robotic solutions that measure current operator's status, adapting the interaction accordingly, and providing her/him with proper training to improve the interaction and learn lacking skills and expertise. In this paper we propose an evaluation and validation procedure to guarantee the achievement of the requirements of a \MATE{} system.
\end{abstract}

\section{Introduction}
Despite the massive introduction, in modern industrial plants, of complex automatic machines and robotic systems, the role of human operators remains crucial. On the one side, automation systems are required to face basic functions with high efficiency, thus allowing higher demands for fast production rate with high quality, and advanced functions, such as fault diagnosis and fast recovery, fine-tuning and reconfiguration of process parameters to adapt to production changes \cite{Sheridan_2002}. On the other side, human operators are responsible for controlling and supervising manufacturing activities and the desired flexible production. However, complex automation solutions typically do not explicitly consider the needs of the human operators: the complexity of the machines is reflected in the complexity of the accompanying human-machine interfaces (HMIs), used for data visualization, for monitoring the processes, and for letting the user interact with the machines~\cite{Skripcak_2013, Nachreiner_2006}.


Typical HMIs generally do not allow to adapt the amount or the form of displayed information. Furthermore, the control systems implemented on industrial machines and processes are generally programmed to respond in a specified way, in order to optimize some performance indexes. Quality and quantity of the information provided to the operator are generally not taken into account, nor adapted to the status of the operator her/himself~\cite{Viano_2000}: the human operator is then, generally, the only element in the production system that is requested to adapt her/his behavior based on the situation. In particular, the operator needs to be able to exhibit a high level of flexibility, to address both common activities and unpredictable situation (e.g., in the presence of alarms or malfunctions). This becomes even more relevant considering the fact that the amount of monitored data that comes from modern production processes is constantly increasing, and control systems are becoming increasingly complex~\cite{Flaspoler_2010, Viano_2000, Skripcak_2013}. 

As a consequence, human operators are exposed to challenging working conditions, from the cognitive and psychological point of view. In particular, while experienced operators are still able to manage the complexity of modern industrial HMIs at the expense of very high levels of cognitive workload, vulnerable operators (such as those with low experience or education level, the elderly and the disabled) experience serious difficulties in the interaction with such systems. 
This may generate several undesirable effects for these operators, ranging from low levels of productivity, to lack of satisfaction for their working condition, to loss of the job (and/or difficulties in re-employment). 

To avoid these effects, a possible approach is to reverse the design of complex production systems and adopt a responsible approach based on the anthropocentric design methodology. This
consists in a user centered design process that ensures that the needs of workers and operators are met, the resulting system is understandable and usable, it accomplishes the desired tasks, and the experience of use is positive and enjoyable~\cite{Norman_2013, ISO9241}. In the context of industrial production, this amounts to reverse the paradigm from the current belief that ``the human learns how the machine works'' to the future scenario in which ``the machine adapts to human capabilities'' accommodating to her/his own time and features~\cite{Skripcak_2013,Nachreiner_2006}. 

This scenario falls within the concept of context-dependent automation, also known as adaptive automation~\cite{Parasuraman_2000, Lee_2013}. Generally speaking, context awareness is the ability for a system to sense, interpret, respond and act based on the context~\cite{Dey_2001}. Based on this paradigm, the level of automation of a system is designed to be variable, depending on situational demands during operational use. Including feedback from the user into the definition of the context leads to modulating the interaction on the basis of her/his current psychophysical status. Specifically, we refer to the need to take into account the physical and cognitive overload induced by the working task and use such information to adapt the interaction accordingly, with the ultimate aim of relieving user's stress and affect.

Along these lines, a few works have recently appeared that propose adaptation methodologies, for reducing the amount of stress on the operators interacting with complex systems. The idea is that of changing how the information is presented, in such a way that only the relevant pieces of information are provided to the operator, based on the context. These concepts have been applied to different application domains, such as automotive \cite{Sharma_2008, Amditis_2006, Garzon_2012}, aeronautics \cite{Inagaki_2000} and smartphones and hand-held devices \cite{Gu_2004}.

However, to the best of the authors' knowledge, only a few pioneering examples have been preliminarily presented regarding HMIs for complex industrial systems~\cite{Viano_2000, Lee_2013}. Specifically,~\cite{Viano_2000} describes a preliminary concept of architecture for an HMI that adapts the presentation of information based on the operator responsiveness. Profiling of the operators is considered in~\cite{Lee_2013}, and the HMI selectively presents information based on the profile of the current user.


Recently, a new methodological approach was proposed in~\cite{Villani_2018_RAM} that is referred to as \MATE{}, i.e., \emph{M}easure, \emph{A}dapt and \emph{TE}ach. This approach builds upon the paradigm of affective computing and robotics, which rely on  measuring user's physiological parameters that are known to be related to mental strain, and adapting the interaction with the automated system accordingly~\cite{Picard_1997, Landi_2017_MECH}. More specifically, the \MATE{} approach consists in merging the concepts of anthropocentric design of human-machine/robot systems and of affective computing: not only is current operator's status measured to adapt the interaction accordingly, but also she/he is properly trained to permanently improve the interaction and learn lacking skills and expertise. In this framework, a thorough characterization of the worker that includes skills, perceptive and cognitive capabilities in addition to mood and affect, and thus extends measurements proper of affective computing, allows a more tailored adaptation of the interaction tasks and training support. 

The contribution of this paper is in the definition of procedures for the evaluation and validation of a \MATE{} system, in order to verify that it is an effective approach to guide and support human operators during human-machine/robot interaction. In particular, procedures will be defined for testing the behavior of each component independently, and for evaluating the system as a whole, thus assessing the satisfaction of the requirements that characterize a \MATE{} system. Indeed, as discussed in Section~\ref{sec:architecture}, the design of a \MATE{} system builds upon a set of application independent requirements that guide the design of such an adaptive interaction system. In this paper we discuss how to evaluate the design of a \MATE{} system with respect to such requirements and identify which tests and measurements are needed.

The paper is organized as follows. Section~\ref{sec:architecture} describes the architecture of a \MATE{} system, its main components and the main requirements to be satisfied. Section~\ref{sec:component} introduces procedures for the evaluation of each component of a \MATE{} system independently. Section~\ref{sec:system} defines procedures for the evaluation of the system as a whole, to verify the satisfaction of the main requirements. Finally, Section~\ref{sec:conclusions} contains some concluding remarks.

\section{\MATE{} system architecture}\label{sec:architecture}
\begin{figure}
	\centering
	\includegraphics[width=\columnwidth]{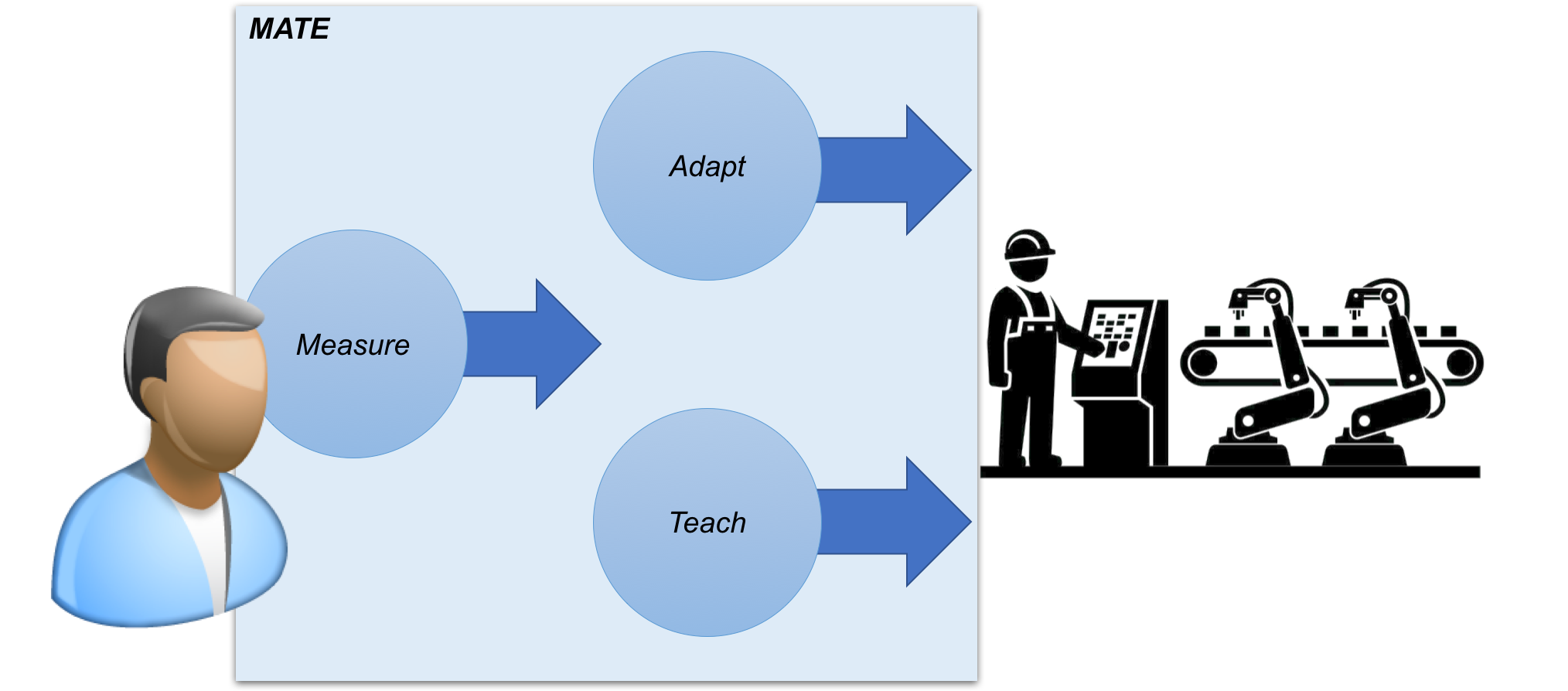}
	\caption{\label{fig:overview_inclusive}Overview of a \MATE{} system.}
\end{figure}


The concept of \MATE{} system was introduced in~\cite{Villani_2018_RAM}, and consists in a smart interaction system that adapts the quantity and quality of the information in the HMI, as well as the functionalities of the machine, to the  physical, sensorial and cognitive capabilities of workers. 

Three main pillars, as depicted in Fig.~\ref{fig:overview_inclusive}, constitute a \MATE{} system:
\begin{enumerate}
	\item  Human capabilities measurement (\emph{Measure}): the system measures the human capability of understanding the logical organization of information and the cognitive burden the operator can sustain (automatic human profiling). Moreover, the system identifies the actual skill level of the user, analyzing on-line how she/he operates in the common working processes.
	\item Adaptation of interfaces to human capabilities (\emph{Adapt}): the system adapts the organization of the information (e.g. the complexity of the information presented), the means of interaction (e.g. textual information, only graphics, speech, etc.), and the automation task (normal operation, adaptation to new processes, predictive maintenance, etc.) that are accessible by the user depending on her/his measured capabilities.
	\item Teaching and training for unskilled users (\emph{Teach}): a virtual environment can be used to provide initial training based on a simulation. The system furthermore teaches the correct way to carry out a task to the unskilled user. Depending on the skill level of the user and the operation performed by the machine, the interface guides the user by means of a step by step procedure.
\end{enumerate}


\begin{table*}
	\centering
	\caption{\label{tab:design_recommendations}Design recommendations for a \MATE{} system, as introduced in~\cite{Villani_2018_RAM}: technical, ELSI, and roboethics requirements.}
	\includegraphics[width=1.8\columnwidth]{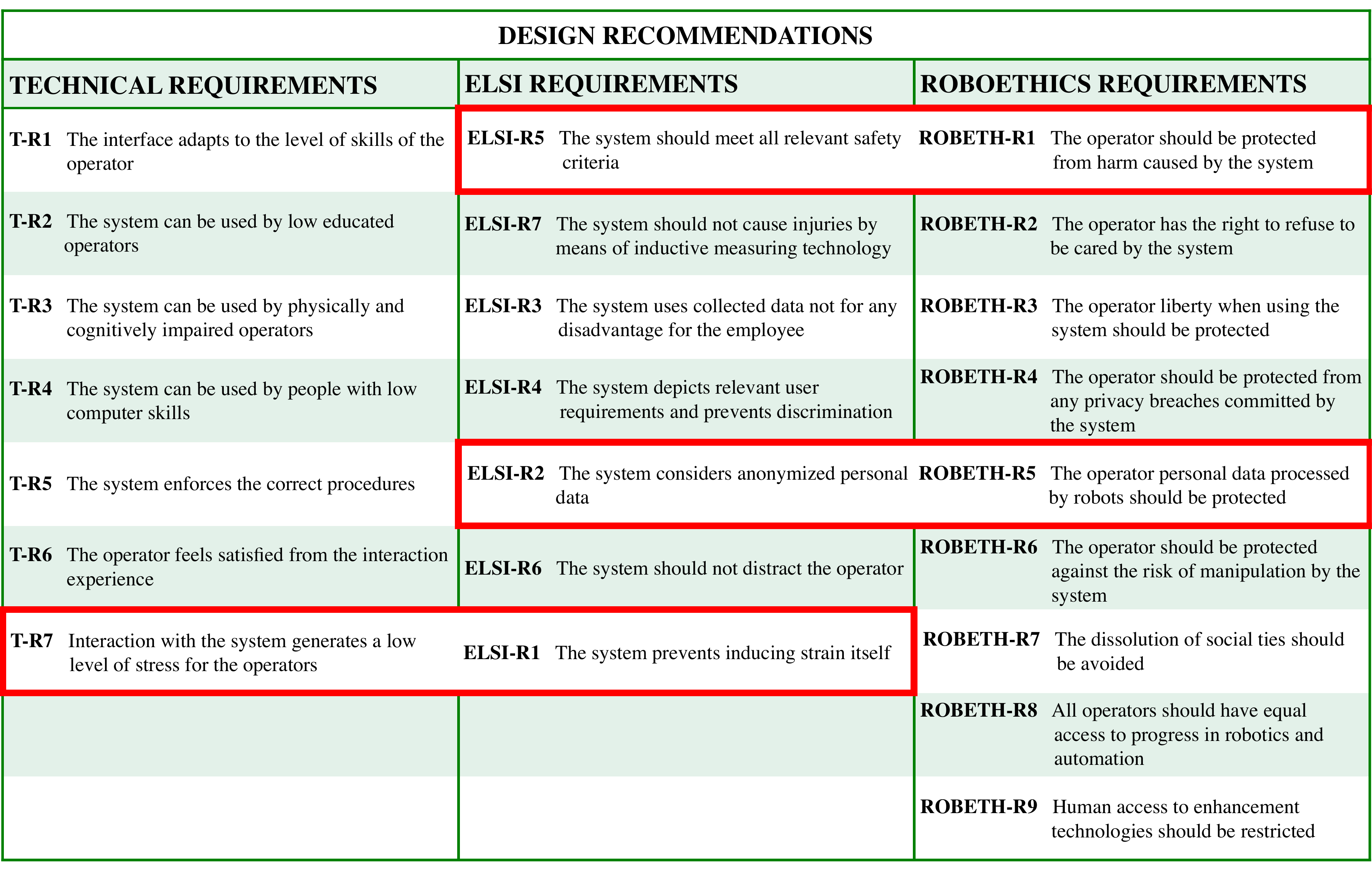}
\end{table*}

A general methodology for the design of \MATE{} systems has been proposed in~\cite{Villani_2018_RAM}. Therein, the general requirements that need to be fulfilled by a \MATE{} system have been discussed, and they are summarized in Table~\ref{tab:design_recommendations}. While system specifications depend on the specific application under consideration, the requirements can be defined from a universal point of view, for \MATE{} systems in general. Such requirements are derived based on an anthropocentric approach, focusing on the users' needs. They represent application independent design recommendations that, on the one side, define what should be done from a technical point of view to ensure proper adaptation to the user and, on the other side, highlight how this should be done in a manner that preserves the user her/himself.

More specifically, technical requirements were first considered, and were subsequently complemented by additional requirements, as well necessary, regarding Ethical, Social and Legal Implications (ELSI) of the system. Finally, roboethics requirements~\cite{Veruggio_2016} were considered. These were introduced in the literature to address the ethical issues related to the use of robots. However, they apply also to the case of production machinery. 


\section{Evaluation procedure at component level}\label{sec:component}
%
As discussed in Section~\ref{sec:architecture}, a \MATE{} system is composed of three main modules, namely \emph{Measure}, \emph{Adapt} and \emph{Teach}. While the effectiveness of each module in satisfying the requirements needs to be evaluated considering the system as a whole, the functionality of each module needs to be verified by means of appropriate evaluation procedures. 

In the following, we detail the evaluation procedures for the three components of a \MATE{} system. For each module, the number and the characteristics of the test subjects need to be selected, based on the specific target application, in order to guarantee statistical significance of the results.

\subsection{Measure}
The main functionality that the \emph{Measure} module needs to provide is the ability to acquire and elaborate measurements of human capabilities, detecting the correct user profile. More specifically, the set of user profiles of interest depends on the specific application: in general, profiles are defined in terms of working experience, demographic data, and strain level.

Measurements can be grouped in three main classes:
\begin{enumerate}
	\item A priori measurements are performed offline, and consist of questionnaires regarding demographic, cultural and ethnographic information, tests for perceptive, cognitive and physical disabilities, and questionnaires regarding skills and working experience.
	\item Real-time physiological measurements consist in measuring physiological indicators for mental strain, such as pupil diameter, blinking rate, skin conductance, cerebral activity, body temperature, hormonal balance and heart rate.
	\item Real-time performance measurements consist in tracking performance indicators, such as time for decisions, execution steps for a task, number of mistakes or redundancies.
\end{enumerate}

A priori measurements can be performed utilizing well known tools, such as the \emph{Vienna Test System} for cognitive diagnostics and the \emph{Motoric performance series} for physical capabilities, which have been extensively shown to provide reliable measures of perceptive, cognitive and physical capabilities.

Real-time physiological measurements are well known in the literature to be related to people's emotional and cognitive condition (see, e.g., \cite{Matthews_2015, Marinescu_2017}). In the application scenario of \MATE{} systems, these measurements must be performed with portable unobtrusive devices that do not limit the freedom of movement of users, are robust to motion artefact and can detect physiological changes as fast as possible. To assess the performance of the \emph{Measure} module, mental fatigue can be induced on purpose by means of stressors commonly known in the literature, such as arithmetic tasks, with penalties on elapsed time and competing players \cite{Ahern_1979, Humphrey_1994}. Tests aim at evaluating whether the \emph{Measure} module can detect incipient conditions of stress and what resolution in stress levels can be achieved.

Regarding real-time performance measurements, it is necessary to assess the ability of the module to identify the correct user profile.  For this purpose, evaluation tests need to be carried out considering subjects with known profile, which can be estimated from the working experience or technical acquaintance, and assessing whether the performed measurements successfully recognize the correct profile. 

\subsection{Adapt}
The main functionality that the \emph{Adapt} module needs to provide is the ability of the HMI to autonomously adapt according to the user profile. Specifically, adaptation rules are devised to adapt the interaction to the user's measured perception and cognition capabilities, and to provide the most suited interaction technology according to the user's characteristics. 

Given a set of adaptation rules, evaluation tests need then to be performed to assess whether they correctly match the needs of the subjects belonging to different user profiles. During the design and implementation phase (i.e., when the final HMI is not yet available), such tests can be performed exploiting mockups of the HMIs, that replicate the expected behavior of the system (or of a portion of the system). 

\subsection{Teach}
The main functionality that the \emph{Teach} module needs to provide is the ability to provide initial training to unexperienced operators and to support the operator during the interaction with the system when needed. 

The teach functionalities can be provided based on two main submodules:
\begin{enumerate}
	\item The off-line training system is used to train unexperienced operators, in a simulated environment, to instruct them on how to interact with the real machine.
	\item The on-line training system supports the operator during the interaction with the machine, providing guidance and instructions.
\end{enumerate}

The effectiveness of the off-line training system needs to be evaluated against a baseline system that does not provide adaptation to the capabilities and the state of the user. The performance of test subjects can be measured by comparing values like the numbers of errors or the execution time.

The effectiveness of the on-line training system needs to be evaluated assessing the performance in the execution of a task. In particular, the experiment needs to be designed to deliberately cause errors in the execution, or high strain levels in the test subjects. Performance in the task execution needs then to be measured, to evaluate the effectiveness of the proposed teaching strategies. 


\section{Evaluation procedure at system level}\label{sec:system}
%
%

\begin{table*}
	\centering
	\caption{\label{tab:measurements}Measurements for the evaluation of a \MATE{} system, to verify technical, ELSI, and roboethics requirements introduced in Table~\ref{tab:design_recommendations}.}
	\includegraphics[width=1.8\columnwidth]{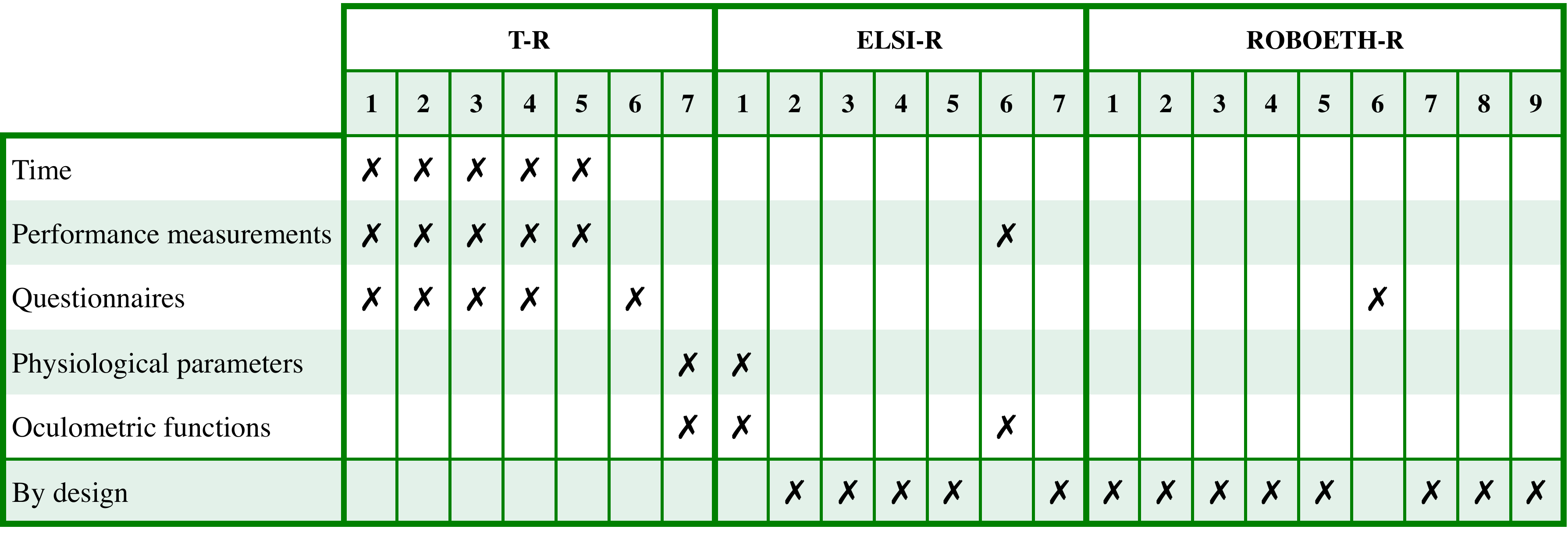}
\end{table*}

The evaluation procedure at system level aims at evaluating the \MATE{} system as a whole, assessing the satisfaction of the requirements discussed in Section~\ref{sec:architecture}. In particular, the goal of these tests is that of analyzing whether the \MATE{} approach effectively alleviates the burden of interaction tasks, increasing job satisfaction and working performance, without causing discomfort to the user, mainly generated by the fact that her/his performance is tracked.

Starting from the list of requirements reported in Table~\ref{tab:design_recommendations}, a set of measurements is derived, to be performed on the overall system. For this purpose, aggregated clusters of requirements are considered, to derive the necessary measurements. In particular:
\begin{itemize}
	\item Requirements T-R1, T-R2, T-R3, T-R4: these requirements are satisfied if the system can be effectively utilized by users with different levels of skills, education, and impairments. This is assessed considering both the performance of the operators, and their satisfaction. As such, it is necessary to measure the \emph{time} required to perform each operation with the \MATE{} system (compared to a baseline system that does not provide any support or adaptation to the user), as well as the \emph{real-time performance measurements} (i.e., tracking performance indicators, such as time for decisions, execution steps for a task, number of mistakes or redundancies). Subjective assessment related to usability of the \MATE{} system is then measured by means of \emph{questionnaires}.
	\item Requirement T-R5: this requirement is satisfied if the system ensures that the operator follows the correct procedures. This is assessed considering the performances in the execution of each operation, in terms of \emph{time} and \emph{real-time performance measurements}, compared to a baseline system that does not provide any support or adaptation to the user.
	\item Requirements T-R6, ROBOETH-R6: these requirements quantify the satisfaction of the operator in the interaction with the system. This is assessed by means of subjective usability \emph{questionnaires}.
	\item Requirements T-R7, ELSI-R1: these requirements are satisfied if the system does not generate large levels of stress or strain on the operator, during the interaction. This is assessed by means of measurements of \emph{physiological parameters} related to cognitive load (compared to a baseline system that does not provide any support or adaptation to the user), and of \emph{oculometric functions} that are an indicator of cognitive strain.  
	\item Requirement ELSI-R6: this requirement is satisfied if the system does not distract the operator during the interaction. This is addessed by means of measurements of \emph{oculometric functions}, and by means of \emph{real-time performance measurements} related to specific operations. 
\end{itemize}

The remaining requirements listed in Table~\ref{tab:design_recommendations} are satisfied by design in a \MATE{} system, and do not require to perform measurements involving test subjects. More specifically, this holds for:
\begin{itemize}
	\item Requirements ELSI-R2, ELSI-R3, ELSI-R4, ROBOETH-R5, ROBOETH-R7: these requirements are satisfied following standard ethics guidelines and privacy regulation (e.g., EU General Data Protection Regulation (GDPR) 2016/679), such as collecting only anonymous data, and prohibiting access to personal information to any external subject.
	\item Requirements ELSI-R5, ROBOETH-R1: these requirements are satisfied by all the machines and systems installed in industrial environments, since safety regulations are mandatory and, hence, always applied.
	\item Requirement ELSI-R7: this requirement is satisfied performing measurements in a non-invasive manner, e.g. using wearable devices (such as wristbands) or external tracking systems.
	\item Requirements ROBOETH-R2, ROBOETH-R3, ROBOETH-R4, ROBOETH-R8, ROBOETH-R9: these requirements are satisfied imposing a correct policy of use for the \MATE{} system. For each application scenario, such a policy should be devised to avoid that the use of a \MATE{} system, and in general any advanced interaction system, becomes harmful to human rights and well being.
\end{itemize}

The previously defined set of measurements to be performed is summarized in Table~\ref{tab:measurements}, compared to the requirements of a \MATE{} system listed in Table~\ref{tab:design_recommendations}.

\subsection{Evaluation scenarios}
Considering the specific target application, \emph{evaluation scenarios} need to be devised, in which such measurements need to be carried out. Evaluation scenarios represent prototypical situations, that are representative of common activities to be performed with the \MATE{} system. While the exact definition of such scenarios depends on the specific application, the following high-level classes of evaluation scenarios are introduced. Table~\ref{tab:ESmeas} summarizes the main measurements that apply to each evaluation scenario.

\begin{table}
	\centering
	\caption{\label{tab:ESmeas}Measurements for the evaluation of a \MATE{} system applied to each evaluation scenario.}
	\includegraphics[width=0.7\columnwidth]{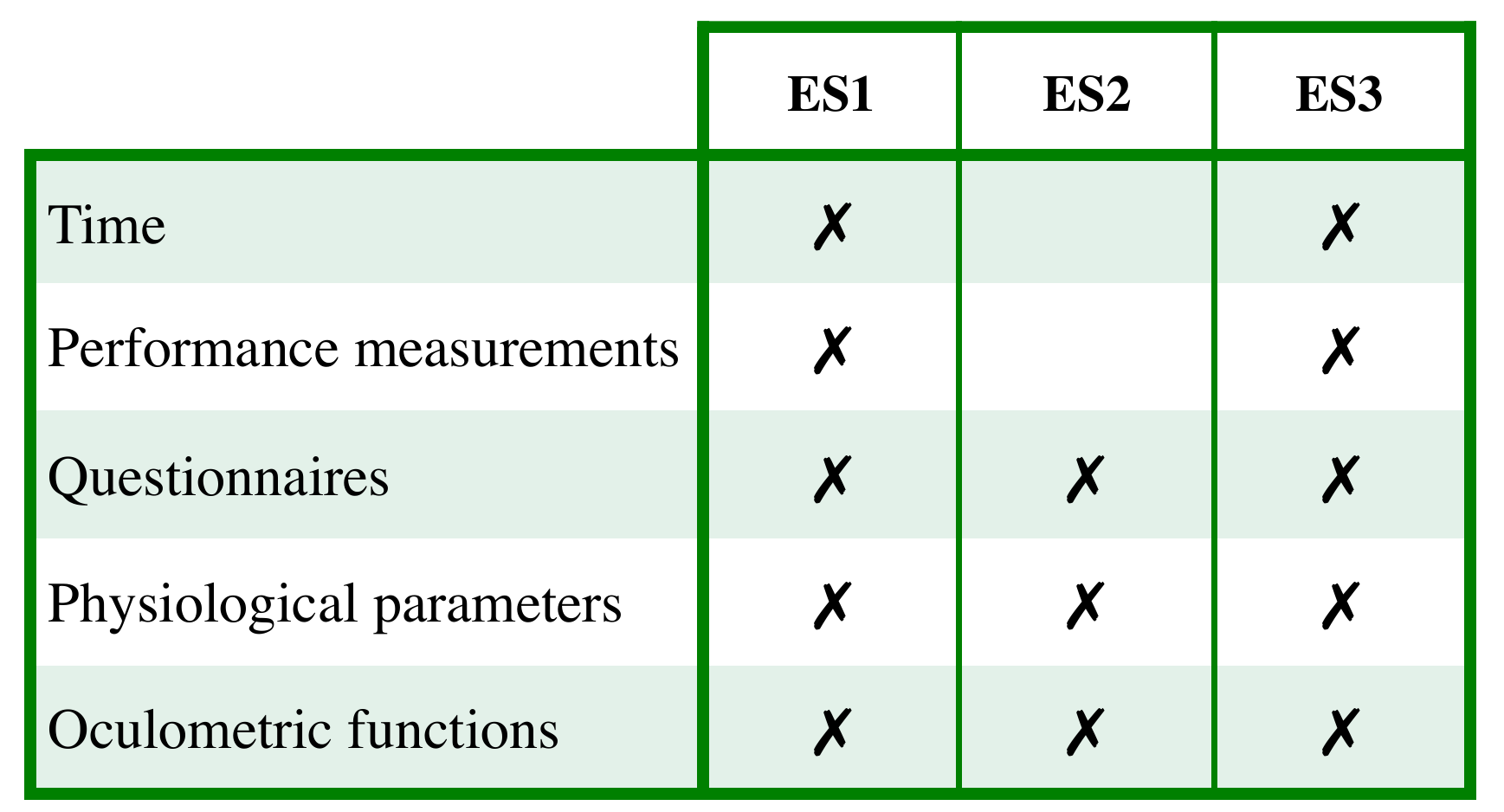}
\end{table}

\subsubsection*{\textbf{ES1 - Procedural}}
A \emph{procedural} evaluation scenario consists in a task for which the user is required to complete an ordered sequence of activities. Examples of such tasks include set-up operations, reconfiguration operations, or ordinary maintenance procedures.

These tasks are characterized by (often very long) lists of activities to be sequentially completed by the user. Besides assessment of usability and induced strain levels, performance metrics can be measured such as time needed to complete the procedure, or number of errors made during the procedure.

\subsubsection*{\textbf{ES2 - Supervision}}
\emph{Supervision} evaluation scenarios consist in supervising complex, and possibly large, systems, possibly including more than one machine. In these scenarios, even considering highly automated solutions, human supervision is generally necessary, to assess, monitor, and control the overall status of the system.

Given the complexity of the system, the amount of information to be considered, simultaneously, is very large: it is then very relevant to assess the level of strain and cognitive load induced on the operator.

\subsubsection*{\textbf{ES3 - Extraordinary maintenance}}
An \emph{extraordinary maintenance} evaluation scenario consists in procedures to be performed in the presence of unpredictable and/or infrequent situations, such as replacing broken or malfunctioning parts, or recovering from anomalies or alarm situations.

Since these tasks are not frequently executed, it can be difficult, for the user, to correctly know, remember and apply the desired sequence of actions. Besides assessment of usability and induced strain levels, performance metrics can be measured, such as time needed to complete the procedure, or number of errors made during the procedure.

\section{Conclusion}\label{sec:conclusions}
In this paper we proposed a novel procedure for the evaluation and validation of a \MATE{} system. First introduced in~\cite{Villani_2018_RAM}, the concept of \MATE{} systems consists in a methodological approach for the anthropocentric design of assistive human-machine systems. In particular, \MATE{} systems are composed of three main modules, which allow to measure the user's capabilities, to adapt accordingly the user interface, and to provide guidance when needed.

The procedure described in this paper provides a set of guidelines to assess the effectiveness of a \MATE{} system in guiding and supporting the human operator during interaction tasks. Evaluation and validation are performed on two levels: at component level (where the functionalities of each module are tested independently) and at system level (where the performance of the overall system is evaluated). Generalized evaluation scenarios are proposed and discussed, and a set of measurements is derived, which do not depend on the specific application domain. 

The methodology proposed in this paper is then general and application independent: the general methodologies proposed here can then be instantiated based on the specific application scenario, complementing the tests that are generally completed when designing and implementing a new system, to compare its performance with technical specifications.

The proposed methodological approach will be applied on the adaptive human-machine systems that will be developed considering real industrial use cases within the EU H2020 INCLUSIVE project~\cite{Villani_2017_ETFA}. The application of the procedure to real-world scenarios will allow to refine both the design recommendations for \MATE{} systems~\cite{Villani_2018_RAM} and the evaluation procedure proposed in this paper. 



\bibliographystyle{IEEEtran}
\bibliography{INCLUSIVE}

\begin{thebibliography}{10}
\providecommand{\url}[1]{#1}
\csname url@rmstyle\endcsname
\providecommand{\newblock}{\relax}
\providecommand{\bibinfo}[2]{#2}
\providecommand\BIBentrySTDinterwordspacing{\spaceskip=0pt\relax}
\providecommand\BIBentryALTinterwordstretchfactor{4}
\providecommand\BIBentryALTinterwordspacing{\spaceskip=\fontdimen2\font plus
\BIBentryALTinterwordstretchfactor\fontdimen3\font minus
  \fontdimen4\font\relax}
\providecommand\BIBforeignlanguage[2]{{%
\expandafter\ifx\csname l@#1\endcsname\relax
\typeout{** WARNING: IEEEtran.bst: No hyphenation pattern has been}%
\typeout{** loaded for the language `#1'. Using the pattern for}%
\typeout{** the default language instead.}%
\else
\language=\csname l@#1\endcsname
\fi
#2}}

\bibitem{Villani_2018_RAM}
V.~Villani, L.~Sabattini, J.~N. Czerniak, A.~Mertens, and C.~Fantuzzi, ``{MATE}
  robots simplifying my work: benefits and socio-ethical implications,''
  \emph{{IEEE} Robot. Automat. Mag.}, vol.~25, no.~1, pp. 37--45, March 2018.

\bibitem{Sheridan_2002}
T.~B. Sheridan, \emph{Humans and Automation: System Design and Research
  Issues}.\hskip 1em plus 0.5em minus 0.4em\relax New York, NY, USA: John Wiley
  \&amp; Sons, Inc., 2002.

\bibitem{Skripcak_2013}
T.~Skripcak, P.~Tanuska, U.~Konrad, and N.~Schmeisser, ``Toward nonconventional
  human-machine interfaces for supervisory plant process monitoring,''
  \emph{{IEEE} Trans. Human-Machine Systems}, vol.~43, no.~5, pp. 437--450,
  Sep. 2013.

\bibitem{Nachreiner_2006}
F.~Nachreiner, P.~Nickel, and I.~Meyer, ``Human factors in process control
  systems: The design of human--machine interfaces,'' \emph{Safety Science},
  vol.~44, pp. 5--26, 2006.

\bibitem{Viano_2000}
G.~Viano, A.~Parodi, J.~Alty, C.~Khalil, I.~Angulo, D.~Biglino, M.~Crampes,
  C.~Vaudry, V.~Daurensan, and P.~Lachaud, ``Adaptive user interface for
  process control based on multi-agent approach,'' in \emph{Proc. Working Conf.
  Advanced Visual Interfaces}, ser. AVI '00.\hskip 1em plus 0.5em minus
  0.4em\relax New York, NY, USA: ACM, 2000, pp. 201--204.

\bibitem{Flaspoler_2010}
E.~Flasp{\"o}ler~et al., ``The human machine interface as an emerging risk,''
  European Agency for Safety and Health at Work, Tech. Rep., 2010.

\bibitem{Norman_2013}
D.~A. Norman, \emph{The design of everyday things: Revised and expanded
  edition}.\hskip 1em plus 0.5em minus 0.4em\relax Basic books, 2013.

\bibitem{ISO9241}
``{ISO} 9241-210: 2010. ergonomics of human system interaction-part 210:
  Human-centred design for interactive systems,'' International Standardization
  Organization {(ISO)}, Tech. Rep., 2009.

\bibitem{Parasuraman_2000}
R.~Parasuraman, T.~Sheridan, and C.~Wickens, ``A model for types and levels of
  human interaction with automation,,'' \emph{IEEE Trans. Systems, Man, and
  Cybernetics - Part A: Systems and Humans}, vol.~30, no.~3, pp. 286--297, May
  2000.

\bibitem{Lee_2013}
A.~Lee and J.~Martinez~Lastra, ``Enhancement of industrial monitoring systems
  by utilizing context awareness,'' in \emph{IEEE Int. Multi-Disciplinary Conf.
  Cognitive Methods in Situation Awareness and Decision Support {(CogSIMA
  2013)}}, 2013.

\bibitem{Dey_2001}
A.~K. Dey, ``Understanding and using context,'' \emph{J. Personal and
  Ubiquitous Computing}, vol.~5, no.~1, pp. 4--7, Feb. 2001.

\bibitem{Sharma_2008}
H.~Sharma, R.~Kuvedu-Libla, and A.~Ramani, ``{ConFra}: A context aware human
  machine interface framework for in-vehicle infotainment applications,'' in
  \emph{Int. MultiConf. Engineers and Computer Scientists {(IMECS 2008)}},
  2008, pp. 19--21.

\bibitem{Amditis_2006}
A.~Amditis, H.~Kussmann, A.~Polychronopoulos, J.~Engstr{\"o}m, and L.~Andreone,
  ``System architecture for integrated adaptive hmi solutions,'' in
  \emph{Intelligent Vehicle Symp.}, 2006, pp. 388--393.

\bibitem{Garzon_2012}
S.~Garzon and M.~Poguntke, ``The personal adaptive in-car {HMI}: Integration of
  external applications for personalized use,'' in \emph{Advances in User
  Modeling}, S.~B. Heidelberg, Ed., 2012, pp. 35--46.

\bibitem{Inagaki_2000}
T.~Inagaki, ``Situation-adaptive autonomy: Dynamic trading of authority between
  human and automation,'' in \emph{Proc. Human Factors and Ergonomics Society
  Annual Meeting}, vol.~44, Jul. 2000, pp. 13--16.

\bibitem{Gu_2004}
T.~Gu, H.~Pung, and D.~Zhang, ``A middleware for building context-aware mobile
  services,'' in \emph{{IEEE} Vehicular Technology Conf. {(VTC 2004)}}, 2004,
  pp. 2656--2660.

\bibitem{Picard_1997}
R.~W. Picard, \emph{Affective computing}.\hskip 1em plus 0.5em minus
  0.4em\relax {MIT} Press Cambridge, 1997, vol. 252.

\bibitem{Landi_2017_MECH}
C.~Talignani~Landi, V.~Villani, F.~Ferraguti, L.~Sabattini, C.~Secchi, and
  C.~Fantuzzi, ``Relieving operators' workload: Towards affective robotics in
  industrial scenarios,'' \emph{Mechatronics}, vol. submitted, 2018.

\bibitem{Veruggio_2016}
G.~Veruggio, F.~Operto, and G.~Bekey, ``Roboethics: Social and ethical
  implications,'' in \emph{Springer handbook of robotics}.\hskip 1em plus 0.5em
  minus 0.4em\relax Springer, 2016, pp. 2135--2160.

\bibitem{Matthews_2015}
G.~Matthews, L.~E. Reinerman-Jones, D.~J. Barber, and J.~Abich~IV, ``The
  psychometrics of mental workload: multiple measures are sensitive but
  divergent,'' \emph{Human Factors}, vol.~57, no.~1, pp. 125--143, 2015.

\bibitem{Marinescu_2017}
A.~C. Marinescu, S.~Sharples, A.~C. Ritchie, T.~S{\'a}nchez~L{\'o}pez,
  M.~McDowell, and H.~P. Morvan, ``Physiological parameter response to
  variation of mental workload,'' \emph{Human factors}, 2017.

\bibitem{Ahern_1979}
S.~Ahern and J.~Beatty, ``Pupillary responses during information processing
  vary with scholastic aptitude test scores,'' \emph{Science}, vol. 205, no.
  4412, pp. 1289--1292, 1979.

\bibitem{Humphrey_1994}
D.~G. Humphrey and A.~F. Kramer, ``Toward a psychophysiological assessment of
  dynamic changes in mental workload,'' \emph{Human Factors}, vol.~36, no.~1,
  pp. 3--26, 1994.

\bibitem{Villani_2017_ETFA}
V.~Villani, L.~Sabattini, J.~N. Czerniak, A.~Mertens, B.~Vogel-Heuser, and
  C.~Fantuzzi, ``Towards modern inclusive factories: A methodology for the
  development of smart adaptive human-machine interfaces,'' in \emph{22nd
  {IEEE} Int. Conf. Emerging Technologies And Factory Automation
  {(ETFA)}}.\hskip 1em plus 0.5em minus 0.4em\relax IEEE, 2017.

\end{thebibliography}

\end{document}